# Modeling membrane curvature generation using mechanics and machine learning

S. A. Malingen and P. Rangamani*

Department of Mechanical and Aerospace Engineering,
University of California San Diego, La Jolla CA 92093.
*To whom correspondence must be addressed: prangamani@ucsd.edu

## Abstract

The deformation of cellular membranes regulates trafficking processes, such as exocytosis and endocytosis. Classically, the Helfrich continuum model is used to characterize the forces and mechanical parameters that cells tune to accomplish membrane shape changes. While this classical model effectively captures curvature generation, one of the core challenges in using it to approximate a biological process is selecting a set of mechanical parameters (including bending modulus and membrane tension) from a large set of reasonable values. We used the Helfrich model to generate a large synthetic dataset from a random sampling of realistic mechanical parameters and used this dataset to train machine learning models. These models produced promising results, accurately classifying model behavior and predicting membrane shape from mechanical parameters. We also note emerging methods in machine learning that can leverage the physical insight of the Helfrich model to improve performance and draw greater insight into how cells control membrane shape change.

## 1 Introduction

Membranes compartmentalize cells while allowing controlled interactions across their interfaces. One of the membrane's core functions is facilitating communication across compartments which can occur by uptake and release (endo- and exocytosis, respectively, when occurring across the external cell membrane). In addition to shaping a cell's microenvironment, these mechanisms act as homeostatic regulators (*1*, *2*). Many elegant models of membrane deformation have been developed (as reviewed in (*3*)), and, in particular, Helfrich's continuum model based on thin shell elastic theory constrained by minimizing bending energy (*4*) is widely applicable. Using the Helfrich model the equilibrium shape of the membrane can be predicted from a set of mechanical parameters (such as the bending rigidity of the membrane, its tension, and many more). Modelers choose mechanical parameters to approximate biological/molecular mechanisms and to match experimentally measured values. Ultimately, this model can be used to show how cells may tune their mechanics to achieve experimentally observed shapes. Furthermore, the Helfrich model can also be used in reverse to determine the forces needed to maintain an experimentally observed membrane shape at equilibrium (*5*, *6*).

One of the primary challenges in each of these applications is prescribing the mechanical parameters of the membrane. Some parameters (such as tension, bending rigidity, pressure and size) have been measured experimentally, so modelers can reference from a rich body of literature to determine reasonable ranges of values. Other parameters, like curvature and the functions used to describe how these values vary over the surface of the membrane, are heuristics that describe an amalgamation of factors. Ultimately, understanding the behavior of the Helfrich model over large parameter spaces is an ongoing challenge, leading to large amounts of labor in hand tuning mechanical parameters to obtain convergent, biologically relevant results. Previously sensitivity analysis has been used to determine how the prescription of the curvature function within the Helfrich model drives membrane energy and shape (*7*), which was a particularly vital advance since preferred membrane curvature is not experimentally accessible. However, to the authors' knowledge the behavior of the Helfrich model over large parameter spaces has not been documented. To fill this gap we present several machine learning models that enable systematic exploration of mechanical parameter space to understand membrane shape changes. In particular, we have used the formation of microparticles (MPs) as a case study for how different biophysical parameters drive shape change.

## 1.1 Microparticle formation is a mechanochemical process

MPs are a type of exocytic vesicle bounded by (and budding from) the plasma membrane in a process reminiscent of blebbing. They provide a mechanism for long-range communication between cells and tissues (*8*) that is vital in health, facilitating functions like blood clotting, but are deleterious in multiple diseases (*9*, *10*). For instance, in cancer, they act as drivers of niche establishment (*11–14*). Increased MP production also occurs in SCUBA divers during decompression (*15*, *16*). In contrast to these examples of MPs in illness, physical exercise can modulate MP production as a signaling component in beneficial vascular adaptations that enhance blood flow to muscles (*17*). Ultimately these are a few examples of how MPs enable long-range communication within an organism, using both their internal contents and the membrane itself as signaling platforms (*9*, *18*). Finally, MP release can be a controlled process; in addition to their functional roles in health and disease, the regulation of their formation is an area of active investigation.

The formation of MPs is an inherently mechanochemical process (*10*). We use spatially varying parameters in our continuum model to approximate the core biological mechanisms observed in MP formation. One of MP formation's hallmarks is the exposure of a negatively charged aminophospholipid, phosphatidylserine (PS), on the external membrane leaflet (*9*). At high concentrations, PS induces membrane curvature (*20–22*). We use a preferred curvature term to account for this process. Curvature can arise from myriad mechanisms (*23*), including from lateral heterogeneity in membrane composition (*11*, *24*), steric interactions between proteins adjacent to the membrane (*25*) (notably, cargo can be localized to the site of MP formation (*11*)), and even the glycocalyx can induce a preferred membrane curvature (*26*). A high cytosolic calcium level in stimulated cells triggers PS exposure (*9*, *27–29*) and can activate calpain, which drives cortical cleavage and cytoskeletal remodeling (*30*). Loss of cytoskeletal integrity can cause a decrease in membrane rigidity (*29*), and, along with cleaving the cytoskeleton, calpain can disrupt linkages between the membrane and the cortex (*31*). The loss of integration with the cortex could allow the membrane to be pushed outwards due to internal hydrostatic force, similar to blebbing. Accordingly, we include spatially varying bending rigidity and outward pressure as tunable parameters in our biophysical model. Finally, all of these parameters coalesce to produce microparticles that span a range of sizes, and so we vary the patch area over which we adjust these parameters. While these biological events are well-documented, it is unclear how they come together mechanically to accomplish MP formation.

We are informed by a rich body of membrane modeling literature (*3*), and our work builds from a previous model of clathrin-mediated endocytosis (*32*), a process that has been extensively studied from both an experimental and theoretical lens (*6*). To adapt our model to MP formation we are inspired by models of blebbing, where the cell membrane also baloons outward (*33–35*). From this model, we derive energetically favorable membrane shapes for randomly sampled parameter combinations from a large mechanical parameter space. Subsequently, we used these data to train machine learning models to predict membrane shape from mechanical parameters. Using this approach we develop a predictive toolbox for mapping a wide range of mechanical parameters accessible in cells to shape outcomes.

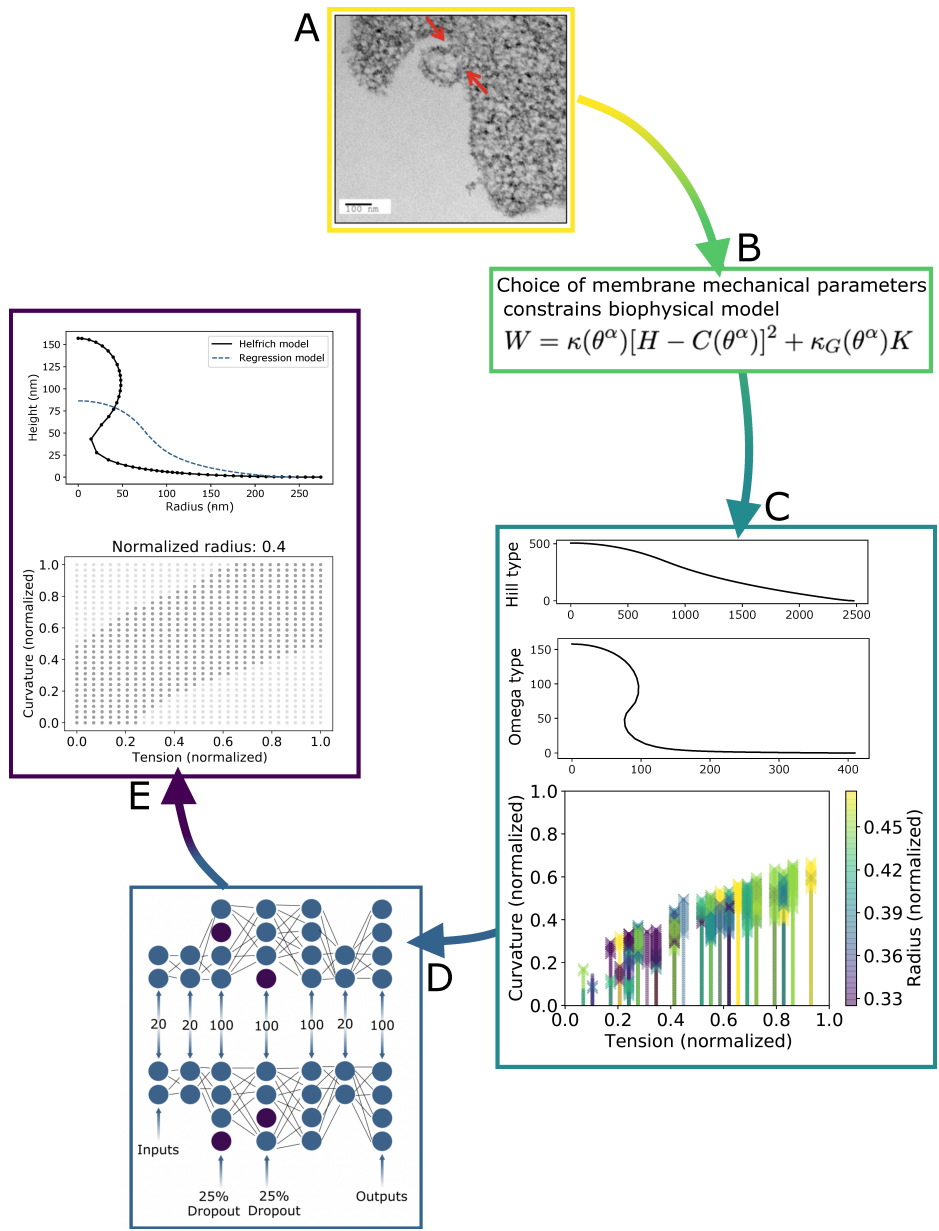

**Figure 1: In this study we used the classical Helfrich model to generate synthetic data that we used to train a machine learning model to predict how the membrane forms MPs given a set of mechanical parameters.** A) Membrane shape changes have been documented with microscopy techniques, for instance during microparticle formation (image from reference (*19*), permission pending). B) Biophysical models, like the Helfrich continuum model, can be paired with best guess mechanical parameters in order to approximate experimentally observed shapes. C) We used the Helfrich model to create a large synthetic data set of membrane shapes and classified the shape as either a hill or omega shaped curve based on if the angle of a tangent to the curve exceeded $90°$. D) Using this synthetic dataset we trained machine learning models to classify model behavior over large parameter spaces, and to predict shapes from mechanical parameters.

## 2 Methods

### 2.1 Synthetic data generation

Ultimately the biophysical model we have used stems from minimizing the energetic cost of deforming a thin 2D membrane (which is in mechanical equilibrium) embedded in 3D space. The Helfrich energy functional describes the energetic cost of bending the membrane (*4*). As was done in reference (*32*), we have used the Helfrich energy functional modified to allow spatially varying preferred curvature as a constitutive equation for the membrane:

$$W = \kappa(\theta^\alpha)[H - C(\theta^\alpha)]^2 + \kappa_G(\theta^\alpha)K$$

where $\kappa$ is the bending rigidity of the membrane which can vary spatially over the coordinate system $\theta^\alpha$, where $\alpha \in \{1, 2\}$; $H = \frac{1}{2}(k_1 + k_2)$ is the mean curvature of the membrane with $k_1$ and $k_2$ denoting the principal curvatures of the membrane; and $C$ describes the spontaneous curvature of the membrane, which can vary spatially like the bending rigidity. Similarly, $\kappa_G$ is the Gaussian bending rigidity while $K = k_1 k_2$ is the Gaussian curvature.

We have adapted the model for membrane bending constrained by the Helfrich energy as derived by Hassinger *et al.* in (*32*). This is valid since for both inward and outward budding the signs of the two principal curvatures match, which means that the Gaussian curvature, $K$, is equivalent for inward and outward budding. And since the signs of the principal curvatures are the same, the absolute magnitude of the mean curvature does not change, although its sign changes. Hence, in our model there is no difference in the energetics of inward versus outward budding. The mathematical underpinnings of this model, and more generally, of thin fluid elastic membrane modeling, are explored in (*36*). In brief, we assume that the membrane is in mechanical equilibrium, so the divergence of the stress vector field of the membrane summed with the pressure applied normal to the membrane is balanced by externally applied forces. This assumption also means that we neglect dynamics. We also assume that the bilayer is incompressible, so a Lagrange multiplier prescribing that the bilayer's density is constant can be imposed. These two observations lead to reconstituting the energy of the membrane in terms of the Helfrich energy summed with the Lagrange multiplier. Finally, the equations of motion can be reframed to yield the force balance normal to the membrane:

$$\begin{aligned} p + 2\lambda H = \Delta[\kappa(H-C)] + 2H\Delta\kappa_G - (\kappa_G)_{;\alpha\beta}b^{\alpha\beta} \\ + 2\kappa(H-C)(2H^2 - \kappa) - 2\kappa H(H-C)^2, \end{aligned} \quad (1)$$

and the tangential force balance within the membrane:

$$\lambda_{,\alpha} = \frac{\delta\kappa}{\delta\theta^\alpha}(H-C)^2 + 2\kappa(H-C)\frac{\delta C}{\delta\theta^\alpha} - \frac{\delta\kappa_G}{\delta\theta^\alpha}K. \quad (2)$$

To simplify finding solutions to these force balances we constrain ourselves to axisymmetric solutions where $s$ describes the position along the arc, $\theta$ describes the rotation of this arc about the $z$ axis, and the $r$ axis describes the radial distance from the $z$ axis. The angle between the tangent ($a_s$) to the arc and the base plane is denoted $\psi$. As demonstrated by Hassinger *et al.* (*32*) a system of 6 ordinary differential equations (ODEs) can be derived from these constraints. To solve this system we used the same 6 boundary conditions, namely at $s = 0^+$:

$$R(0^+) = 0, \ L(0^+) = 0, \ \psi(0^+) = 0, \tag{3}$$

and at $s = S$:

$$Z(S) = 0, \ \psi(S) = 0, \ \lambda(S) = \lambda_0. \tag{4}$$

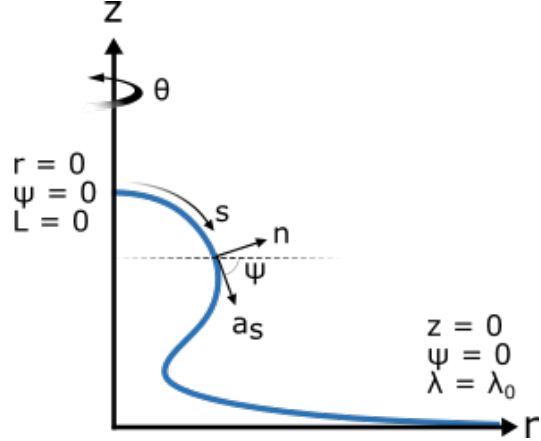

**Figure 2:** Schematic illustrating the key variables used to prescribe the model, and the boundary conditions that constrain it.

These are clamp boundary conditions describing that at the beginning and end of the simulation domain the angle $\psi$ is fixed at $0$. The variable $\lambda$ is interpreted as tension (*37*, *38*), and so while tension can vary along $s$, the tension at the outer edge of the domain is fixed at $\lambda_0$. We used the area parameterization of the system.

## 2.2 Mechanical parameter choice

### 2.2.1 Spatial heterogeneity

The composition of the membrane varies laterally, and we propose that variation in the mechanical features of the membrane are a core component of MP formation. Hence, we define a patch of the membrane at the center of the simulation domain where the mechanical parameters are altered from their values over the rest of the membrane. Since cells can produce MPs spanning a broad range of sizes we explored a range of patch sizes. We aimed to model the formation of microparticles ranging from 25–500 nm in radius ($r$), so we set the patch area to be $4\pi r^2$, while the larger simulation domain was chosen as $4\pi(3r)^2$.

### 2.2.2 Tension

The apparent tension at the cell surface that constrains shape change is the sum of the in-plane tension in the plasma membrane with the adhesion of the membrane to the cortex (*39*). Proteins linking the membrane to the cortex impede the flow of lipids into a tether creating viscous resistance that can be measured with dynamic tether pulling, characterizing the contribution of the cytoskeleton to apparent membrane tension (*40*). Meanwhile, in-plane membrane tension can be characterized by isolating the membrane from the cortex, along with several other methods (*41*). Since microparticle formation requires detaching from the cortex we are most concerned with these values.

By measuring the tension of the membrane in blebs the contribution of cytoskeletal adhesion can be removed, resulting in values near 0.003 pN/nm in rabbit renal proximal tubule epithelial cells and about 0.012 pN/nm in human melanoma cells while the tension in membrane associated with the cortex can reach about 0.022 pN/nm and 0.044 pN/nm in these cell types, respectively (*42*). These values are similar to the roughly 0.003 pN/nm apparent membrane tension measured in neuronal growth cones (*39*), and the roughly ten times larger tensions that have been measured in neutrophils (*43*, *44*).

While we have chosen a relatively small range of tension values, experimental measurements of membrane tension vary broadly with reports ranging from 0.003–0.276 pN/nm, while the in-plane membrane tension has been reported to take values from 0.003–0.15 pN/nm (*41*). We have chosen tension to fall within 0.003–0.15 pN/nm since it encompasses both the reported values for in plane membrane tension, along with the apparent tension for several cell types, but we neglect the upper range of apparent tension values that have been reported. It is noteworthy that tension varies across cell type, state, and across stretching regimes (*41*, *45*), and its role in global and local regulation, and the rate of its propagation through the membrane, is an active area of discovery (*40*, *41*).

### 2.2.3 Bending rigidity

The bending rigidity of the cell membrane is often derived from experimentally measured values in Giant Unilamellar Vesicles (GUVs) (*46*), or red blood cells (*47*) using techniques such as fluctuation analysis (*e.g.* (*48*)), or micropipette aspiration (*e.g.* (*49*)). By using a reduced system, the influence of individual constituents on membrane mechanical properties can be identified. For instance, the inclusion of sterols can increase the bending rigidity of some membranes (*50*), but not all (*46*). While these techniques have provided an extensive literature base to estimate the rigidity of a lipid bilayer (usually between 10–150 $k_B T$ (*47*, *51*)), the bending rigidity of the cell membrane *in vivo* may be heterogenous and take a range of values. Bending rigidity can be influenced by a host of factors, ranging from charge (*46*), to extensions from the membrane like the glycocalyx, protein interactions with the membrane (*52*), or interaction with the cortex (*33*), and likely many other complexities. In light of these challenges, Steinkühler *et al.* have measured the bending rigidity of the membrane in Giant Plasma Membrane Vesicles (GPMVs), effectively maintaining a biological membrane composition while isolating the mechanical properties of a cell membrane from cytoskeletal interactions. Likely, the bending rigidities they measure are similar to that of microparticles, falling roughly in the range of 10–35 $k_B T$ (which at a temperature of 298 K corresponds to roughly 40–145 pN nm). The bending rigidity they measure in isolated membrane derived vesicles is similar to that measured in expanding blebs (*33*, *51*), but when blebs begin retraction the rigidity of the membrane increases to about 215 pN nm, which is attributed to the formation of actin under the lifted membrane (*33*). Further, Charras *et al.* found that treatment with wheat germ agglutinin could increase the membrane rigidity to approximately 360 pN nm. To simplify our model we set a fixed bending rigidity of the membrane surrounding the site of microparticle formation at 320 pN nm. Given the breadth of measured values for membrane rigidity, we allow the rigidity of the region from which the microparticle forms to fall within the range of 40–400 pN nm$^{-1}$. In our model setup we use a multiplier to denote the difference between base membrane rigidity and the rigidity of the patch where a microparticle is expected to form. Based on the range of bending rigidities we found reported in the experimental literature we chose this multiplier to be in the range

0.125–1.25.

### 2.2.4 Pressure

To model hydrostatic pressure we have applied an outward normal force ranging from zero to 0.0003 pN/nm$^2$ to the membrane on the patch region. These values were based on the values reported in a number of studies which were reviewed in reference (*53*), although we chose to explore only the lower range of pressure values they report. As is also reviewed in (*53*), hydraulic pressure is a critical driver of cell blebbing, which bears similarity to MP formation.

### 2.2.5 Curvature

In order to obtain convergent results from the continuum model we initialized a flat membrane with zero mean curvature. Then we looped over small, increasing steps in curvature where the initial guess fed to the model was the curvature at $i-1$ and the model was set to find an energetically favorable solution for the curvature at $i$. For many parameter sets the model did not converge after a given value of curvature. So while curvature values could reach 0.07 nm$^{-1}$ (which corresponds to a radius of about 14 nm), this was rare. Instead, the model was allowed to slowly increase curvature in step sizes of 0.0007 nm$^{-1}$ until it no longer converged.

## 2.3 Data generation

In order to obtain a convergent result from the system of ODEs the solver needs a decent initial guess. Therefore we start with a curvature of zero over the initial patch and a flat membrane as an initial guess and iterate over gradually increasing values of curvature, as described in the preceding methods section dedicated to curvature. Since the Helfrich model produces snapthrough instabilities for some parameter regimes (*32*) we used the final solution that produced a convergent result for increasing values of curvature as an initial guess for a loop where we iterated over decreasing values of curvature. Often, we could not find a convergent solution. While on a case by case basis factors like the solver, mesh density or initial guess could be adjusted to search for a solution this would be infeasible for the range of mechanical parameters we set forth to explore. Hence, the absence of a convergent solution in our data set does not preclude the existence of one. We used *scipy.integrate.solve_bvp* to solve the system of ODEs with the default Runge-Kutta method of order 5(4), a tolerance of 1×10$^{-2}$ and the maximum number of nodes set to 1,000 times the number of points on the mesh (2,000). The mesh itself performed much better when the initial mesh between 0 and 1 was squared so that the density of points about the center (which corresponds to the center of the patch) was larger. The mesh was then resized to the total patch area.

## 2.4 Machine learning model development

In this work we aimed to use machine learning to classify membrane shape with the goal of creating phase maps that delineate regions of parameter space where omega shaped buds are probable. We also created a regression model to predict equilibrium membrane shape explicitly from input parameters.

### 2.4.1 Classification models

We used the criteria that $\psi > 90°$ to label omega shaped buds. In all cases the model was tasked to predict if the membrane shape was an omega shaped bud, or not (usually a hill-shaped or flat bud). The features provided to the model were the physical parameters used in the continuum model of membrane shape. These include: mean curvature, the tension at the membrane's edge, the size of an initial patch or membrane, the bending rigidity of the patch relative to the rest of the membrane, and an outward normal forced applied to the membrane patch which models pressure on a region of the membrane severed from the cortex. Since these parameters alone yield a five dimensional feature space we simplified this problem to two four dimensional parameter spaces: in the results section we first address how curvature, tension, patch size and pressure impact the formation of omega shaped buds, then we address how curvature, tension, patch size and a difference in bending rigidity over the patch region can give rise to omega shaped buds.

While there are many machine learning methods that can be used for this type of classification problem (*54*), we restricted ourselves to xGBoost models and neural net models, creating one of each for each training set. The hyperparameters of the models were tuned by hand, and the model's performance at this stage was determined on a withheld data set composed of 3,741 points for the data set varying pressure, and 2,201 points for the data set varying the difference in bending rigidity of the MP patch from that of the rest of the membrane.

### 2.4.2 Methods for imbalanced classification problem

Both of the data sets are imbalanced, meaning that there are many more non-omega shaped membrane profiles than omega shaped ones. The data set investigating the impact of pressure has a total of 24,499 samples with only 1,513 (about 6.2%) classified as omega shaped buds. The data set investigating bending rigidity has even fewer omega shaped buds, with a total of 18,304 data points and only 453 (about 2.5%) representing omega shaped curves. Since imbalanced classification is an important problem (for instance in medical diagnostics), many approaches have been created to optimize classifier performance. Broadly, these can be viewed as adjusting the ratio of the classes in the training set (for instance, synthetic minority oversampling technique and iterations thereof (*55–58*)) and re-weighting the minority class within the model framework itself. We used two of the simplest approaches: in the xGBoost model we used a built in hyperparameter (scale_pos_weight) to give preference to the minority class (*59*), while in the neural net model we re-sampled the training set to have a ratio of 7:1 majority to minority data points.

### 2.4.3 Neural net

Our neural net model was heavily based on a tutorial from the tensor flow core for classification on imbalanced data (*60*). In brief, we used a sequential model with 5 dense layers and one dropout layer. Dropout is a useful tool for preventing model over-fitting and enhancing a model's ability to generalize for an unseen test set (*61*). We used the rectified linear activation function in all layers except for the final output layer, where we used a sigmoid activation function for binary classification. The model was constructed with 4 hidden layers and a dropout layer, in addition to a single output layer: 10/100/100/50% dropout/10/1. The model architecture and weights are available in the GitHub repository: https://github.com/RangamaniLabUCSD/Modeling_membrane_curvature_generation_using_mechanics_and_machine_learning. We used a batch size of 32.

We used a callback function (see ref: (*62*) for a tutorial on callback functions) for early stopping with a patience of 200 epochs and a maximum of 4,000 epochs was allowed during model training.

| xGBoost hyperparameters | |
| --- | --- |
| objective | binary:logistic |
| learning_rate | .01 |
| max_depth | 100 |
| min_child_weight | 10 |
| gamma | 1 |
| subsample | .2 |
| colsample_bytree | .9 |
| seed | 23 |
| n_estimators | 700 |
| scale_pos_weight | $\frac{\text{\# of non-omegas}}{\text{\# of omegas}}$ in train set |

Table 1: The hyperparameters prescribing the xGBoost model were chosen manually.

Additionally, we used another callback function to save the best model developed during model fitting, also with a patience of 200 epochs. In both cases these callbacks were set to monitor the AUPRC (which is explained in the following section).

### 2.4.4 xGBoost

xGBoost can be a highly effective machine learning method for classification problems (*63*). Selecting hyperparameters to prescribe the model requires balancing a model that can both account for the complexity of the data and mitigate over-fitting. In table 1 we have recorded the hand tuned hyperparameters that we used to define the model. While we chose to hand tune hyperparameters in this instance, the performance of an xGBoost model can be increased through hyperparameter optimization, which is an active area of research (for example, see publications such as (*64–66*)).

### 2.4.5 Measuring classification model performance

Confusion matrices are one way to asses a classification model's performance, allowing the reader to determine the number of times a model correctly predicts each class, and the number of times it missclassified. In a binary classification problem each of the boxes represents either a true negative, false negative, true positive or false positive. However, our classification models do not return strictly binary results; instead they return a probability (between 0 and 1) of an instance belonging to the 0 or 1 class. The confusion matrices presented for each model in this paper represent the model's prediction when a threshold of .5 was chosen for determining class membership. However, this threshold can be increased or decreased, resulting in infinitely many possible confusion matrices for a single trained ML model. To measure performance across all possible thresholds the numbers of a confusion matrix can be embedded in a single metric, and the measure reported along a curve. Then the model's performance over all thresholds can be quantified as the area under the curve. The metric for model performance needs to meet the needs of the classification problem.

Since there is a large imbalance in the number of omega to non-omega membrane shapes in our data set several commonly used metrics of classification performance (namely accuracy and the receiver operating characteristic (ROC)) are inappropriate since they count the number of correctly identified majority class instances. To better quantify the model's capacity to correctly identify the minority class we use the precision-recall curve (PRC). Precision is the proportion of times that the model correctly predicted the minority class (the true positives divided by the total number of positive predictions). The recall (or sensitivity) of the model, in contrast, is the proportion of actual positive cases that the model correctly identifies (the number of true positives divided by the number of false negatives plus true positives). Because neither precision or recall take into account the number of majority cases that the model correctly identifies the PRC highlights the model's ability to discern the minority class without being influenced by the model's ability to accurately predict the majority class. For a helpful tutorial on these ideas see ref: (*67*). Precision and recall values, then, are individual values attributed to a model with a given threshold, so the precision recall curve (PRC) is parameterized by threshold values in the interval [0,1] for a given model. Model performance can then be assessed from the area under the PRC (the AUPRC).

## 2.5 Neural net for regression

Phase-space visualizations compresses the shape of the membrane to a binary indication of behavior for a given set of mechanical parameters. While this reduction is useful, we also sought to create a regression model that used mechanical parameters as the input features to predict the diverse shapes the membrane may adopt. The training data set for this problem was generated in the same step as the regression problem. In addition to whether or not the membrane adopted an omega shape we recorded the x and y locations of 50 points interpolated along the curve (down sampling from an initial mesh of 2,000 points). The neural network was tasked to then predict from a point in mechanical feature space the x and y locations that define the curve. The dense network was composed of a nine layers, two of them dropouts and six dense layers, and finally a dense output layer of 100 nodes, with a structure of 20/20/100/25% dropout/100/25% dropout/100/20/100. We used a rectified linear unit activation function, Adam optimizer with a small learning rate of 0.001, batch size of 64 and mean absolute error (MAE) loss function. Similar to the classification problem we used callbacks for early stopping and a checkpoint to record the model that produced the lowest MAE across all training epochs, both with a patience of 200 epochs. In total, the model was limited to a maximum of 4,000 epochs.

### 2.5.1 Measuring regression model performance

The regression model was tasked to minimize the MAE between individual 'x' and 'y' coordinates along an objective curve. In reporting model performance we again used the MAE and supplemented it with a scaled error metric since large membrane patches with relatively similar shapes can receive a higher MAE than smaller membrane patches with very different shapes. The scaled error was computed by first dividing both the prediction and the actual membrane curve by the maximums of the x and y locations, respectively, of the validation curve. By handling it in this way differences in the height and diameter of the membrane between the prediction and actual data are still apparent, but are scaled relative to true curve's dimensions. Then the minimum Euclidean distance between each point of the predicted curve was computed across all of the points documented along the true curve. Ultimately, this value is still greater than or equal to the distance between the predicted point and the target curve. This second metric is a truer estimate of the model's capacity to correctly predict membrane shape regardless of size. Using a metric like this as the loss function for the model itself, however, is not practical since we seek to predict a relatively even distribution of points along the entire curve while this particular error metric could be minimized by placing all points near the outer clamp boundary condition where the 'y' coordinate approaches zero and the 'x' coordinate could be inferred from initial patch radius.

We inquired if there is a correlation between error and the distance between the test (or validation) point in mechanical parameter space and the nearest training point. Since these parameters are measured in different units and cover unrelated numerical ranges measuring "distance" in mechanical parameter space is necessarily contrived. We normalized the range of each mechanical parameter to fall within [0,1] and, since we used a 4D parameter space, used the L1 (Manhattan) norm to compute the distance (*68*).

# 3   Results and Discussion

Cells deform their membranes in a host of processes, and here, we used MP formation as a case study. In modeling MP formation, we apply the elegant Helfrich continuum model to derive energetically favorable membrane shapes for a given set of mechanical parameters. Shape transformations depend on a host of mechanical properties of the membrane, such as the bending rigidity of the bilayer, its tension, and more; these also constrain biophysical models of membrane shape change. While some of these parameters are experimentally accessible, they can vary spatially and temporally within a cell, making model parameterization non-trivial. Further, the curvature parameter, in particular, is used in continuum models to approximate the effect of an amalgamation of forces (protein-induced spontaneous curvature, steric repulsions, and charge interactions, for instance) that drive the membrane to bend. Therefore the preferred curvature of a membrane is a heuristic that cannot be measured, even while model outcomes depend intimately on its prescription (*7*). Historically, biophysicists have achieved relevant model results by manually tuning the mechanical parameters based on ranges from experimental work and using techniques like grid search to inform phase behavior. However, for high-dimensional parameter spaces, these techniques are very time-consuming and can limit the interpretability of results. Therefore, we sought to use machine learning to represent model behavior over two randomly sampled, 4D mechanical parameter spaces. We present the results of this work into two categories: Classification, where we used xGBoost and a neural network to create phase spaces, and Regression, where we used a neural network to predict membrane shape from mechanical parameters.

## 3.1   Classification

We used the Helfrich continuum model to predict curves representing membrane shape deformations for randomly sampled points in mechanical parameter space. A curve was classified as an omega shaped bud if the angle $\psi$ (Fig 2) exceeded $90°$, and as a hill shaped deformation otherwise. In figures 3 and 4 we show the results of the Helfrich continuum model where parameter combinations yielding omega shaped buds were marked with an 'x', while parameter combinations that yielded a hill shaped curve are marked with dots. The first three rows of these figures show slices of parameter space for small, medium and large ranges of patch area (which corresponds to MP size).

In Fig 3, where pressure was varied, we show that for small and medium patch areas there is a clear trend that as tension increases an increased preferred curvature is needed to drive the formation of omega shaped buds. In contrast, for large patches, only a few parameter combinations resulted in omega-shaped buds, and these usually required relatively high pressures. The final panel of Fig 3 shows a visual separation in the regions of parameter space where omega shaped buds are more likely to occur from regions where they are unlikely to occur. However, the relationship of 4 mechanical parameters is not readily visualized.

The results shown in Fig 4, where the bending rigidity of the patch relative to the rest of the membrane was varied (and applied pressure was set to zero), do not show as clear of a pattern as those in Fig 3 where the pressure applied to the patch area was varied. Generally, omega shaped buds did not occur when the bending rigidity of the patch was much smaller than that of the rest of the membrane. Occasionally, for the largest patch sizes an omega shaped bud could be formed; these tended to occur when the bending rigidity of the patch was larger. Additionally, the same general trend of increasing curvature necessitated with increasing membrane tension was seen, but there was not a clear visual separation of regions of parameter space where MPs may form from where they do not.

These results gave us a visualization for how phase maps should look. Launching from this point we built two machine learning models to predict whether the membrane would adopt an omega shape from the mechanical parameters constraining the Helfrich model. In essence, the models were tasked to interpolate how the membrane would behave based on nearby training points. Both the xGBoost and neural network performed well for the dataset where pressure was varied, with the area under the precision-recall curves (AUPRC) being 0.84 and 0.93, respectively. For the dataset varying pressure, the neural network performed well, with an AUPRC of 0.82 while the xGBoost model had a poorer performance with an AUPRC of 0.39. We chose to measure model performance using the precision-recall curve due to the large class imbalance (with omega shaped buds in the minority class). These models were notably different in two regards. The first is that in the region of parameter space where we were not able to obtain convergent results from the Helfrich model the machine learning models were forced to extrapolate. Interestingly, the neural network guessed in these uncharted regions that the membrane would not adopt an omega shape; meanwhile the xGBoost model guessed that the membrane would adopt an omega shape. Given the limitations of our data sets it is not reasonable to expect either model's predictions to capture actual membrane shape. Further, the absence of data in these regions hints at the possibility that energy-minimizing shapes for these parameter combinations may not exist or different computational schemes may be needed to solve these equations (*7*, *69*, *70*). A second more subtle difference between the predictions of the models is the edges in the phase maps which do not precisely align. It is possible that either model could perform better in these regions under different training conditions such as increasing the sampling of parameter space, particularly around phase transitions.

These classification models may enable researchers to more rapidly tune mechanical parameters by providing a computationally light weight classification. Additionally, they may allow us to bridge the gap between experimental observations of membrane deformation through electron microscopy and computational modeling of membrane mechanics.

## 3.2 Regression

Compressing the membrane shape to a binary categorization of omega or non-omega shaped buds allowed us to create phase diagrams that succinctly convey trends. However, while this compression provides an excellent overview of how a thin elastic membrane behaves, it limits our ability to understand the diversity of shapes the membrane may adopt. Further, obtaining a convergent result from the Helfrich model for a given set of parameters can be quite time consuming, requiring a user to tune the mesh size and step sizes. Therefore we sought to train a regression model to explicitly predict membrane shape for a given set of mechanical parameters. One of the challenges of this problem, in particular, is that conceptually we are not seeking to find the location of individual points but instead to trace a curve. In contrast to this conceptual goal, the regres-sion framework we used is tasked to predict 50 'x' and 'y' locations along the curve, minimizing the distance between the predicted 'x' location and the actual 'x' location, and likewise for the 'y' location. While a custom loss function minimizing the distance to the curve itself could be defined, that would likely result in a model that accurately predicts the most stable portion of the membrane at the outermost edge while neglecting points along the region of the curve that varies the most (the bud region).

Embedded in the goal of predicting a two dimensional curve is the need for a suitable metric of the similarity of the two curves to measure error. We used the MAE (averaged over all of the 'x' and 'y' coordinates) to create a loss function and to measure the model's performance. As we previously noted, this is an overestimate since the predicted point could be closer to the curve than it is to the target point. Additionally, we used the L1 norm, which yields an error estimate greater than (or equal to) the L2 norm. Since the MAE takes size into account, it is possible that large MPs with a close shape match could have a greater error than small MPs where the shapes have a large discrepancy. Therefore we supplemented our measurements of model performance with a second metric that first normalizes size based on the maximum 'x' and 'y' values of the data from the Helfrich model and then uses the L2 norm to calculate the distance to the target point. In designing the ML model, we maintained variation in the size of the membrane patch rather than normalizing and tasking the model to predict shape alone since the mechanics of forming large MPs are different than small ones, as demonstrated in Fig 3 and Fig 4.

In Fig 9B we show the curves in the test set for which the model yielded the highest normalized error and MAE error for both omega shaped and hill shaped curves. We also display in Fig 9C examples of mean error predictions. These curves yield several instructive insights. First, the ML model predicts results that may be unphysical, crossing over into negative 'x' values. The loss function could be further customized to penalize unphysical predictions like this. Next, we see that using a normalized error metric (blue) allowed a discrepancy in the prediction of a small deformation hill curve to be identified, but the loss function of the model does not penalize this highly since the absolute amplitude of the curve is relatively small. Finally, the highest MAE error omega prediction had an especially difficult task – the result of the Helfrich model for this parameter combination appears to be on the way towards a pearling morphology, which we observed very rarely. Likely, there were very few (or no) curves in the training set that resembled this one, and so an ML model with no physical laws governing it could not predict that the membrane would adopt this shape. In Fig 9C, we display the mean error ML predictions for both omega and hill type curves measured with both MAE and normalized error. Hill classed curves were predicted with lower error than omega class curves. Likely, this is due to the imbalance between hill type and omega type curves in the test set. These case studies are born out across the test set, as shown by the wide spread of the omega shaped curves' error distribution compared to the hill classed curves (Fig 9A). Finally, we asked if the error in the ML model's predictions depended on the distance between the test point in mechanical parameter space to the nearest train point. Performing a regression showed that distance did not explain much of the variation in error (MAE error had an $R^2$ of 0.043 and the normalized error had an $R^2$ of 0.022).

Although we must express the limitations of this regression framework, the model generally returned reasonable predictions and is a first step in this arena. Alternative modeling frameworks are being created that will likely lead to improved predictions. For instance, physics informed neural nets (PINNs) use a custom loss function designed from known governing physics to inform if the model's predictions conform to governing physics and to adjust model weights in accordance with decreasing the model's loss (*71*), and are increasingly a tool applied to biological problems (*72*). Another method merging machine learning with mathematical models is differentiable physics (for instance, see (*73*)). In the future these avenues may create a rapid, robust model that links mechanical parameter space to shape changes of the membrane in the forward direction. These methods could take better advantage of modern computing resources and the robust body of existing physical models and numerical methods. Additionally, the large volume of realistic geometries coming online from electron microscopy data may provide training and test data sets (for instance, the Cell Image Library and Janelia). Further, developments in physics informed machine learning increasingly may allow inverse problems and parameter estimation tasks to be tackled. Eventually, these developments may allow the integration of small experimental data sets to infer sets of mechanical parameters that could give rise to observed shapes (see Fig 10).

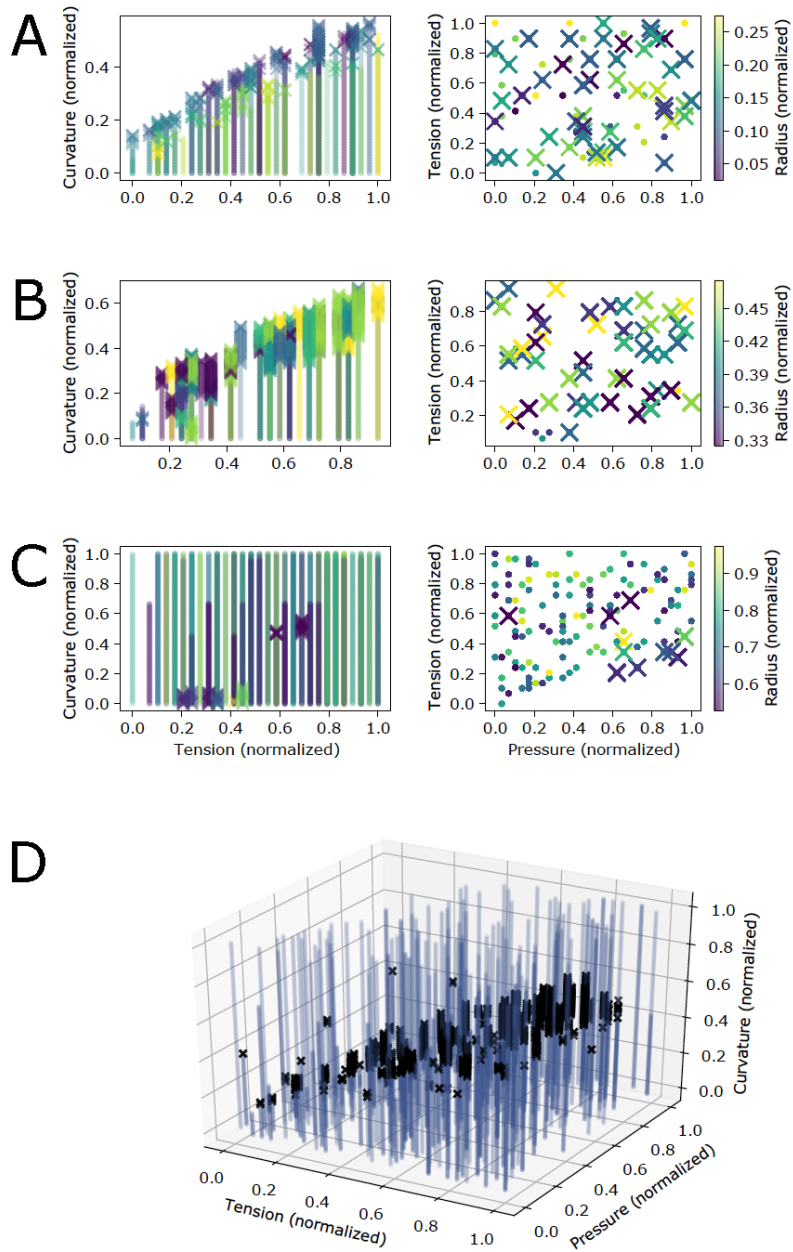

**Figure 3: Results of the continuum model show regions of the 4D mechanical parameter space where omega shaped buds formed.** A) Parameter combinations yielding an omega shaped bud are marked with an 'x' while those yielding a hill shape are marked with a dot. Here we show data for patches of normalized radius smaller than 0.3 with the color map corresponding to the size of the patch. For small patches as tension is increased preferred curvature also needs to increase in order for omega shaped buds to form. B) Similar to small patches, for medium sized patches with normalized radius between 0.3 and 0.5 as tension increases preferred curvature must also increase to yield an omega shaped bud. C) For the largest patches (normalized radius between 0.5 and 1) omega shaped buds are rare, usually only forming for relatively large pressure forces. D) A 3D plot of parameter combinations yielding omega type buds (black 'x's) and combinations yielding hill type deformations (blue dots) shows a band of parameter space where omega type buds are likely to form.

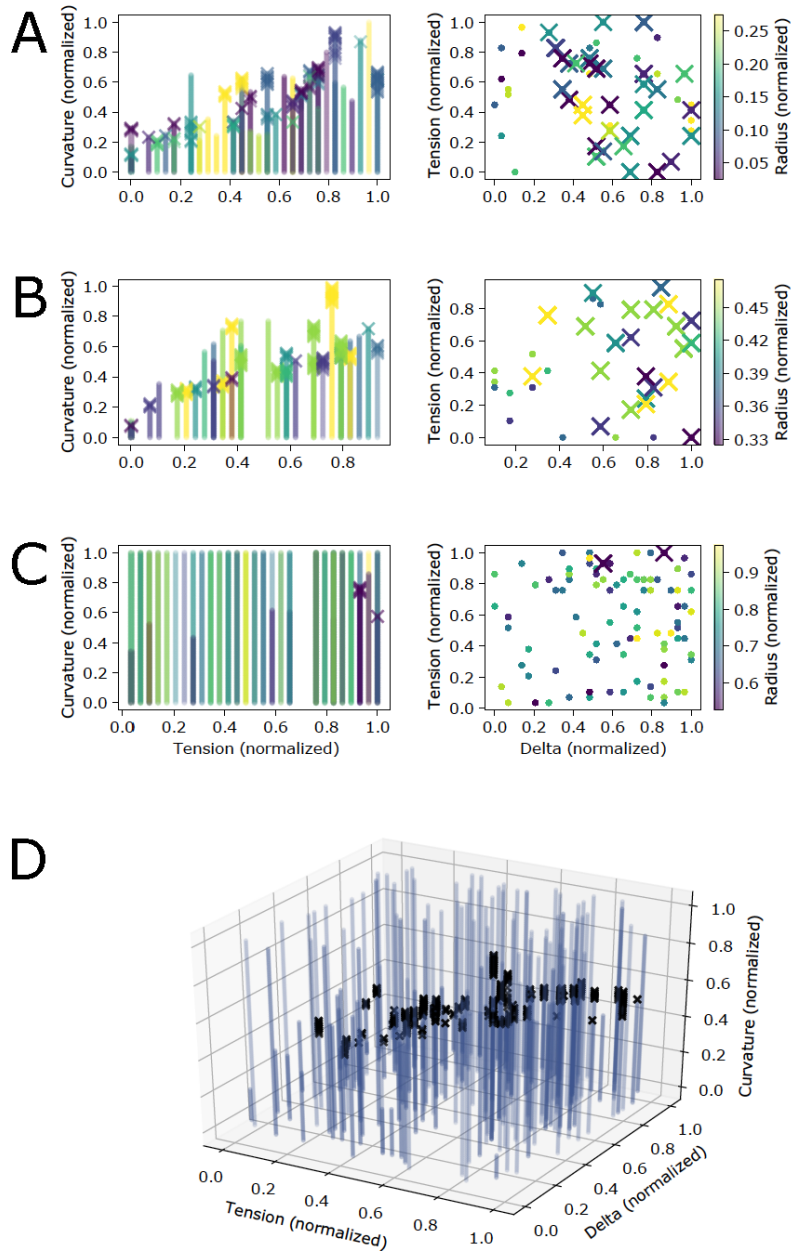

Figure 4: **Change in the bending rigidity of the patch region from that of the rest of the membrane shows that omega shaped buds are less likely to form in regions of parameter space where bending rigidity is lost.** A) Parameter combinations yielding an omega shaped bud are marked with an 'x'. The color map corresponds to the size of the patch. For the smallest patches (normalized radius less than 0.3) omega shaped buds were fairly common except where delta was small (regions where the patch was less rigid than the surrounding membrane). B) A similar trend to small patch sizes was observed for medium sized patches (normalized radius between 0.3 and 0.5). C) For the largest patch sizes very few omega shaped buds were formed. D) A 3D plot of parameter combinations yielding omega shaped buds (marked with black 'x's) and parameter combinations yielding hill type deformations (marked with blue dots) shows that most parameter combinations did not yield omega typed buds.

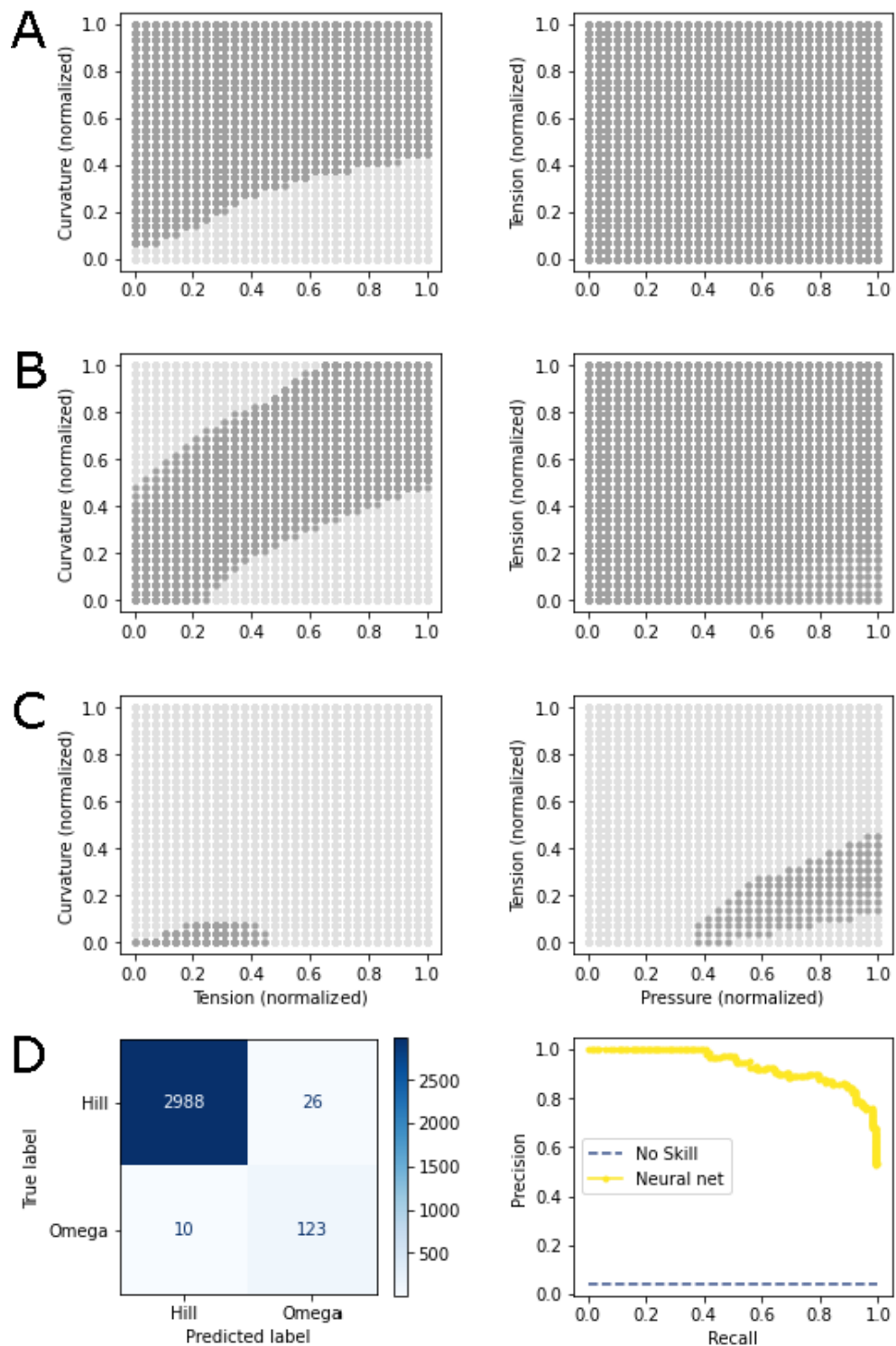

**Figure 5: A neural network yields an accurate phase map of mechanical parameter space.** The model recapitulates the general trends observed visually in Fig 3. A) Model predictions for patches with a normalized radius of 0.15. B) Model predictions for patches with a normalized radius of 0.4. C) Model predictions for patches with a normalized radius of 0.7. D) The confusion matrix for this model with a threshold of .5 shows generally good performance and emphasizes the class imbalance of the data set. The Precision-Recall plot for this model demonstrates strong performance, with an area under the curve of 0.93.

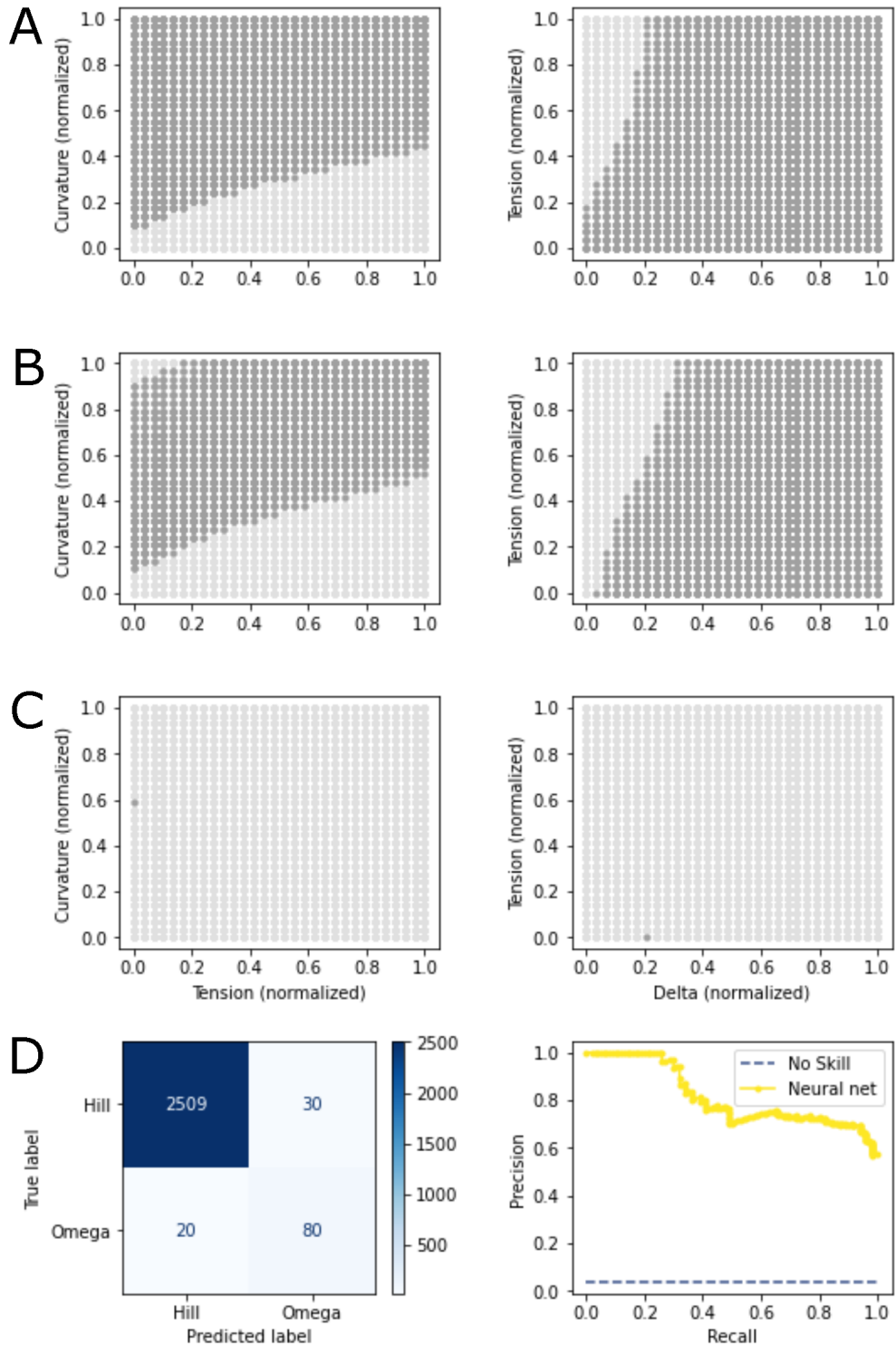

**Figure 6: A neural network model is able to create a phase map of mechanical parameter space.** A) Model predictions for patches with a normalized radius of 0.15. B) Model predictions for patches with a normalized radius of 0.4. C) Model predictions for normalized radius of 0.7. D) The confusion matrix for this model demonstrates that while overall the model was able to correctly classify the small number of positive results it had a small bias to identify false positives. The AUPRC was still relatively high at 0.82.

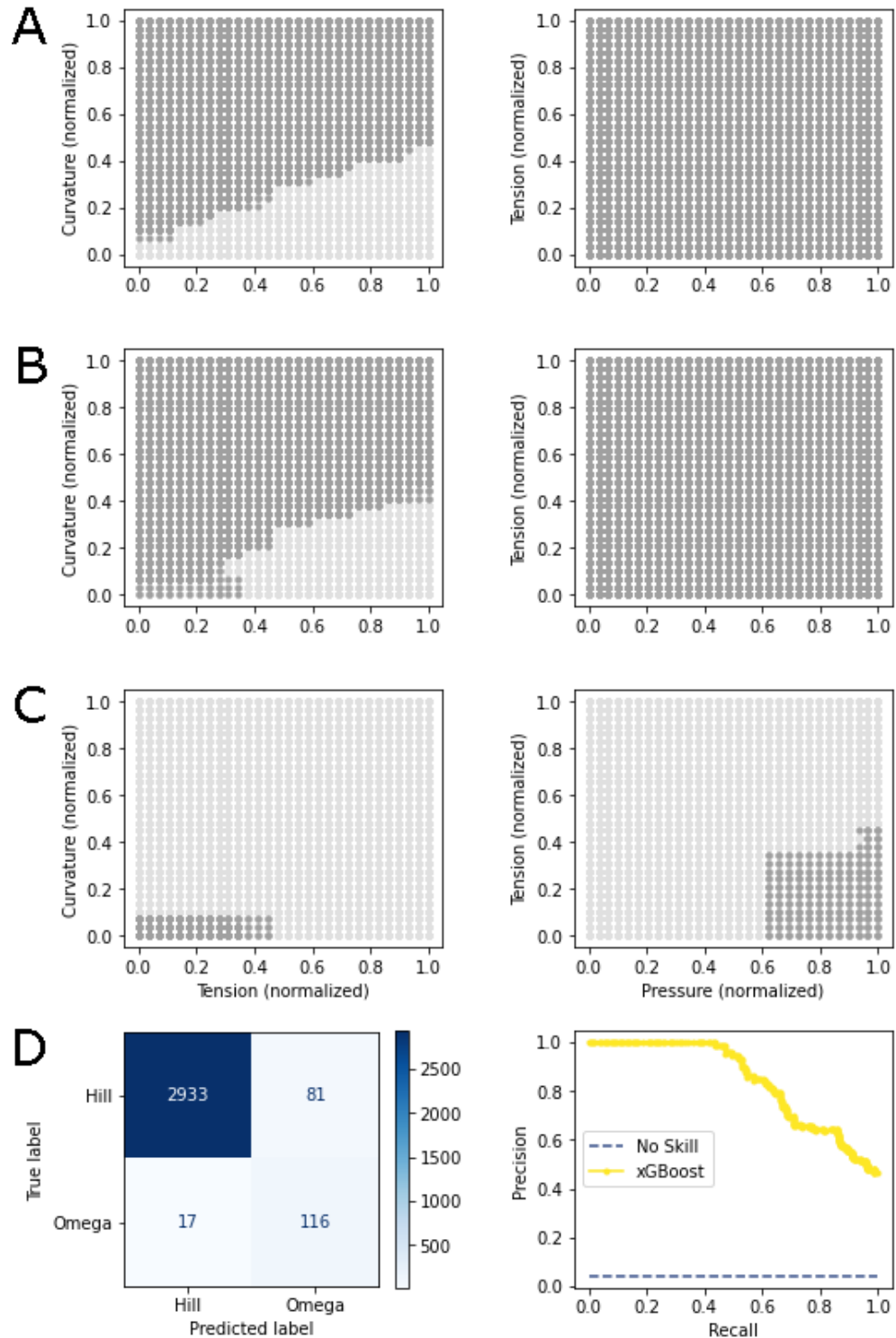

**Figure 7: An xGBoost model captures a phase map of mechanical parameter space**. A) Model predictions for patches with a normalized radius of 0.15. B) Model predictions for patches with a normalized radius of 0.4. C) Model predictions for patches with a normalized radius of 0.7. D) The confusion matrix for this model demonstrates that the model has a relatively high false positive rate. The model overall had high performance with an AUPRC of 0.84.

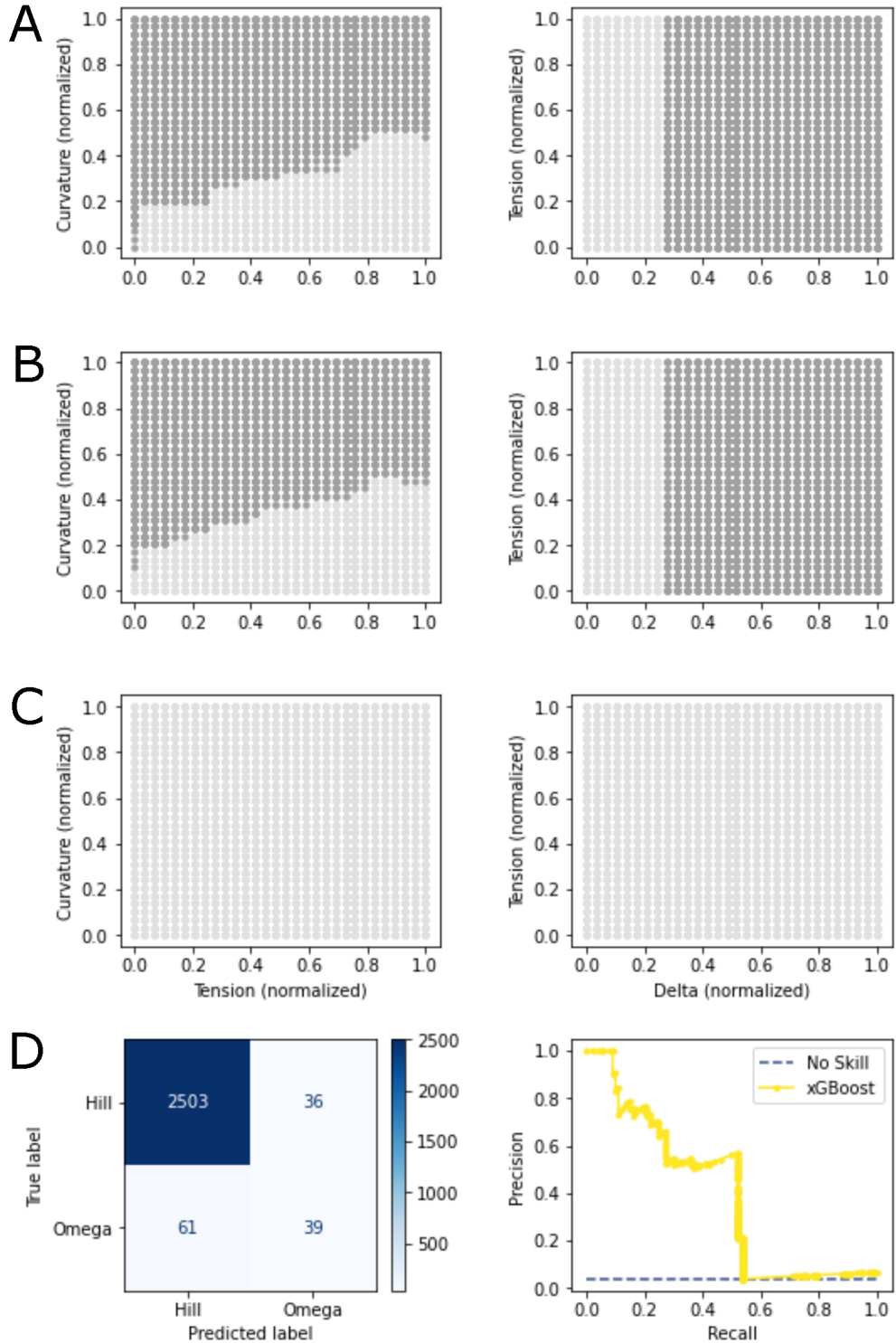

**Figure 8: The xGBoost model makes a reasonable prediction of phase behavior.** A) Model predictions for patches with a normalized radius of 0.15. B) Model predictions for patches with a normalized radius of 0.4. C) Model predictions for patches with a normalized radius of 0.7. D) The confusion matrix for this model shows that the model had a false positive rate near the true positive rate, and a false negative rate larger than the true positive rate. Nonetheless, the model still had an AUPRC of 0.39.

# 4  Conclusions and future directions

In this paper, we developed two machine learning models that demonstrate the feasibility of classifying membrane behavior across mechanical parameter space and predicting membrane shape given mechanical parameters. The parameter spaces we explored are still relatively small (4D). Going forward, if larger parameter spaces are to be explored, care must be taken due to the *Curse of Dimensionality* (for instance, see references (*68*, *74*)), which means that as the number of dimensions grow, random sampling will preferentially explore the edges, rather than the core, of parameter space. A future avenue in need of exploration is to determine which parameters are the most important drivers of shape change in order to guide the selection of a reasonably sized space and parametric sensitivity analysis can help with narrowing down the subset of parameters. We also demonstrate optimistic performance of a deep neural net trained to predict membrane shape from mechanical parameters. ML models like this one could eventually drastically reduce the time needed to visualize the results of the Helfrich model for a choice of parameters within the bounds described by the training dataset. This may benefit researchers since manual parameter tuning to match model results to biological phenomena can be very time consuming. However, before this goal can be realized, lower error predictions would be desirable (particularly for shapes like omega buds that are only sparsely represented in parameter space). Methods such as physics informed neural nets and the merging of differentiable physics with neural nets may enable lower error predictions and harness the robust physical models available for membrane deformation. And, perhaps one of the most exciting open challenges in this field will be to use shape data drawn from experimental data and to reverse compute the sets of mechanical parameters that might give rise to it.

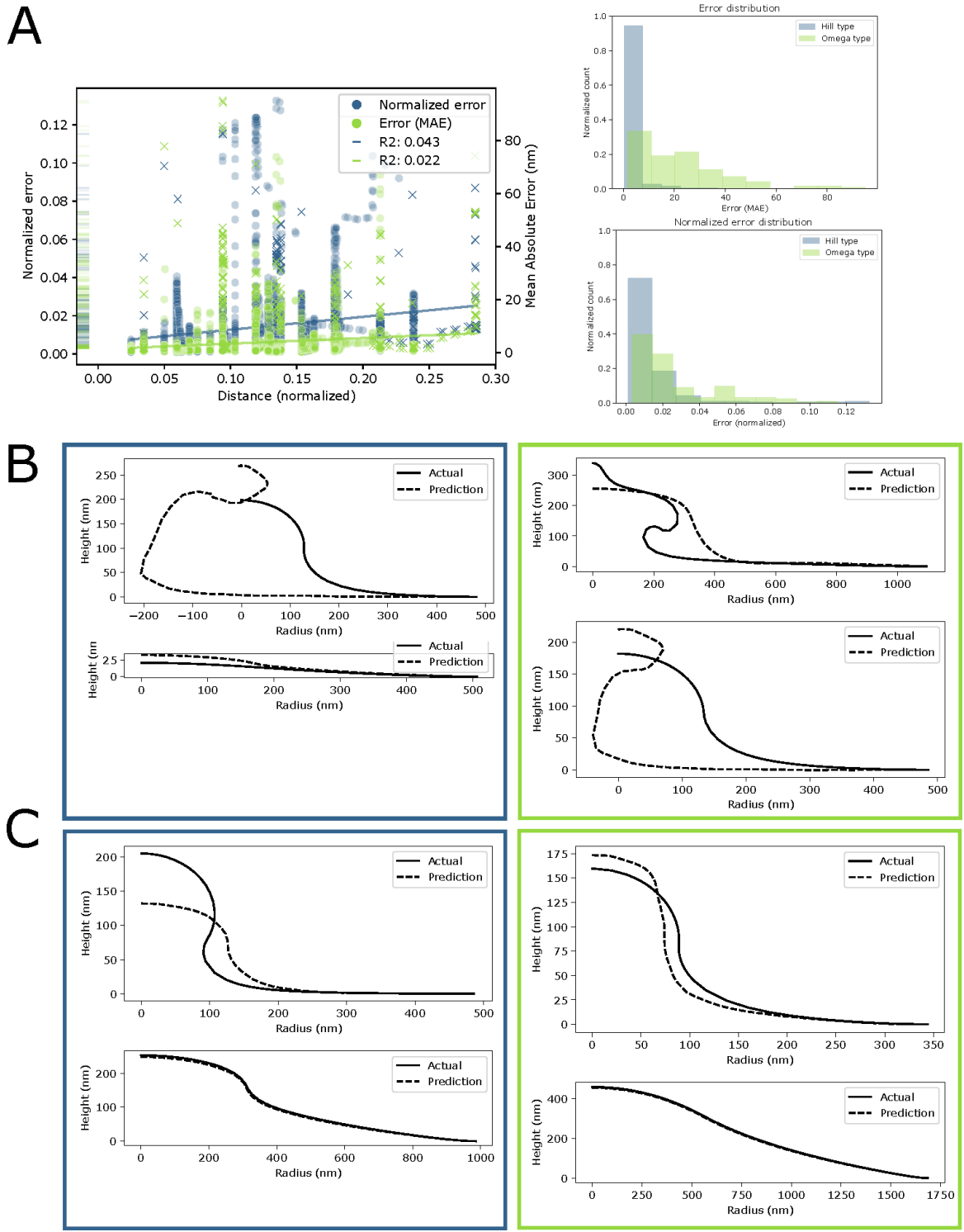

**Figure 9: A neural net model tasked to perform regression to predict membrane shape performs decently well under two error metrics.** A. The error for a single curve is plotted against the normalized distance to the nearest point in mechanical parameter space that was included in the training set. Two error metrics were used – normalized error re-scaled the actual data between zero and one in both the r and z dimensions, and the predicted values were scaled based on the same values as the train data. This metric was used to asses shape change without factoring in scale. Also included is the mean absolute error. Regression on error versus distance shows only very weak correlations between the error of model's predictions and the distance to the nearest training point. The distribution of error is more spread out in MAE for omega shaped buds, but not for normalized error. In panel B) sample of curves that had the largest error predictions for normalized error (blue, left hand column) and MAE (green, right hand column) are shown. In panel C) mean error predictions are shown.

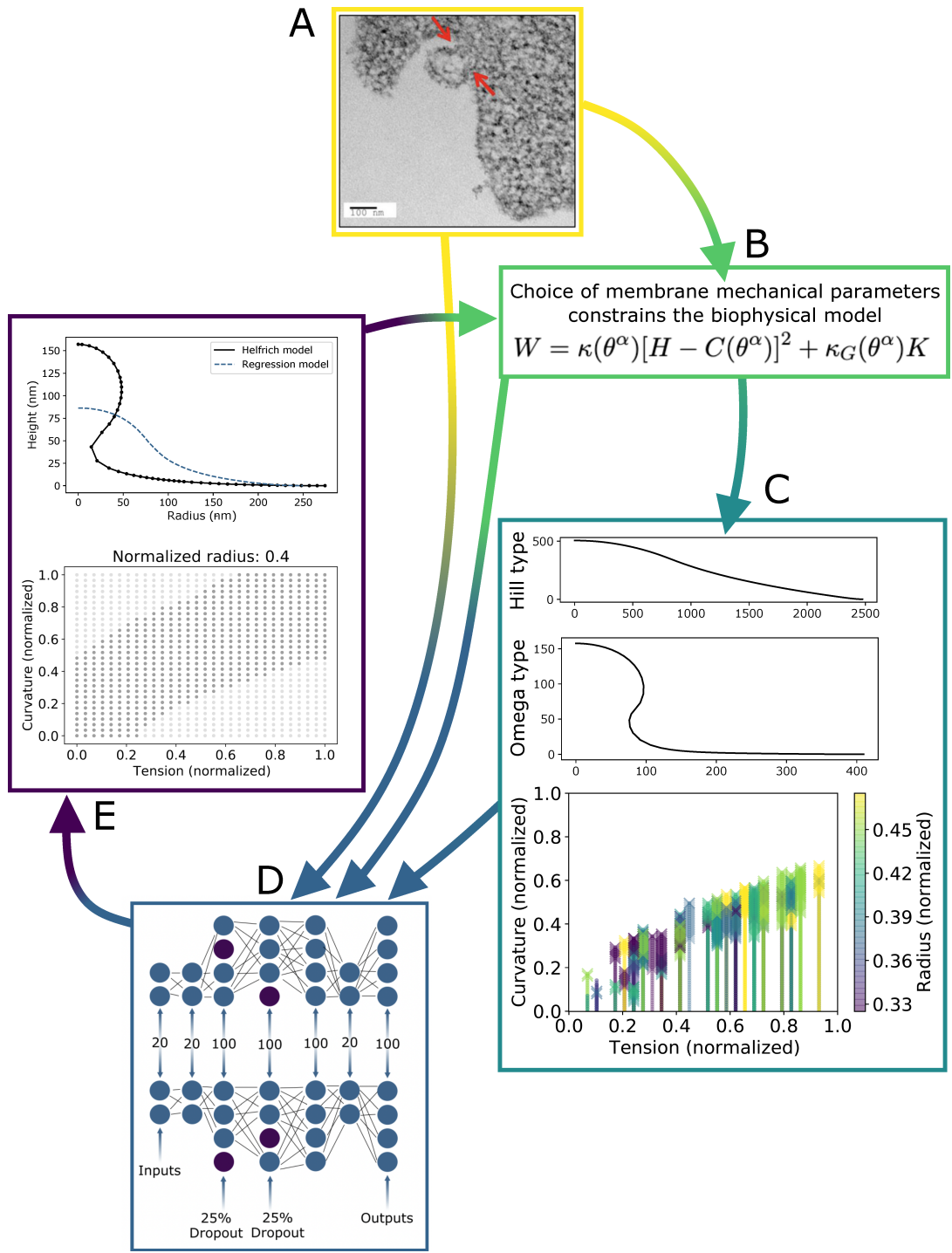

**Figure 10: In the future, emerging techniques in ML could be combined with the Helfrich model to create a more robust and reasoned model, and potentially shed light on how observed membrane shapes relate to the mechanical characteristics of the membrane.** A) In this paper, we did not use an experimentally documented curves (such as those that could be traced from data like that in (*19*)) to train our model, but we suggest that in the future they could be incorporated into a training data set, in addition to inspiring the choice of biophysical modeling constraints. B) The Helfrich continuum model can be used both to generate a synthetic dataset and could also be used to define the error of a machine learning model's predictions. C) Synthetic data can be used in training an ML model. D) ML models that combine powerful physical insight like the Helfrich model with experimental data supplemented with synthetic data may yield more accurate predictions. E) We hope for a future where machine learning models may not only perform forward predictive tasks, but may also use shapes to infer not only the forces needed to sustain them (), but also the sets of mechanical parameters that may give rise to them.

## Acknowledgements

We thank the members of the Rangamani lab who graciously provided their insight into membrane mechanics. SAM was funded by a grant from the Office of Navy Research N00014-20-1-2469 to PR.

## Author contributions

SAM and PR conceptualized the project scope, realized the models in code, and co-wrote the manuscript. Funding was obtained by PR.